\documentclass{eas}
\usepackage{graphicx}
\begin{document}

\title{Radiation Pressure Feedback in Galaxies}
\runningtitle{Radiation Pressure Feedback}
\author{Brett H.~Andrews}
\address{Department of Astronomy, The Ohio State University;
  \email{andrews@astronomy.ohio-state.edu\ \&\ thompson@astronomy.ohio-state.edu}}
\author{Todd A.~Thompson}
\sameaddress{1}
\secondaddress{The Center for Cosmology and Astroparticle Physics, The Ohio
  State University}
\begin{abstract}
We evaluate radiation pressure from starlight on dust as a feedback mechanism in
star-forming galaxies by comparing the luminosity and flux of star-forming
systems to the dust Eddington limit.  The linear $L_{\rm FIR}$--$L_{\rm
HCN}^\prime$ correlation provides evidence that galaxies may be regulated by
radiation pressure feedback.  We show that star-forming galaxies approach but do
not dramatically exceed Eddington, but many systems are significantly below
Eddington, perhaps due to the ``intermittency'' of star formation.  Better
constraints on the dust-to-gas ratio and the CO- and HCN-to-H$_2$ conversion
factors are needed to make a definitive assessment of radiation pressure as a
feedback mechanism.
\end{abstract}
\maketitle
Observations show that the star formation efficiency per free fall time is only
$\sim$1\% (Kennicutt \cite{k98}), likely caused by the injection of
energy/momentum into the ISM by massive stars (``feedback'').  The radiation
pressure associated with the absorption/scattering of UV/optical light from
massive stars by dust grains has been suggested as the dominant feedback
mechanism in star-forming galaxies (Thompson \etal \/ \cite{tqm05} [T05];
Krumholz \& Matzner \cite{km09}; Murray \etal \/ \cite{mqt10} [M10]; Andrews \&
Thompson \cite{at11} [AT]).  If radiation pressure on dust dominates feedback,
then galaxies and their star-forming subregions should approach the dust
Eddington luminosity: $L_{\rm Edd} = (4 \pi G c M_{\rm g}) / \kappa_{\rm F}$,
where $M_{\rm g}$ is the gas mass in the region of interest and $\kappa_{\rm F}$
is the flux-mean opacity, which depends strongly on the column density of the
medium [T05], varying from $\sim$10$^3$ cm$^2$/g in regions that are marginally
optically thin to UV light to a constant value of $\sim$few--10 cm$^2$/g in
regions that are optically-thick to the re-radiated FIR.  In the latter, the
$\kappa_{\rm F}$ depends linearly on the dust-to-gas ratio since more dust
implies a higher efficiency of momentum coupling to the gas.

\begin{figure}
\begin{minipage}[t]{6cm}
\begin{center}
\includegraphics[width=6cm]{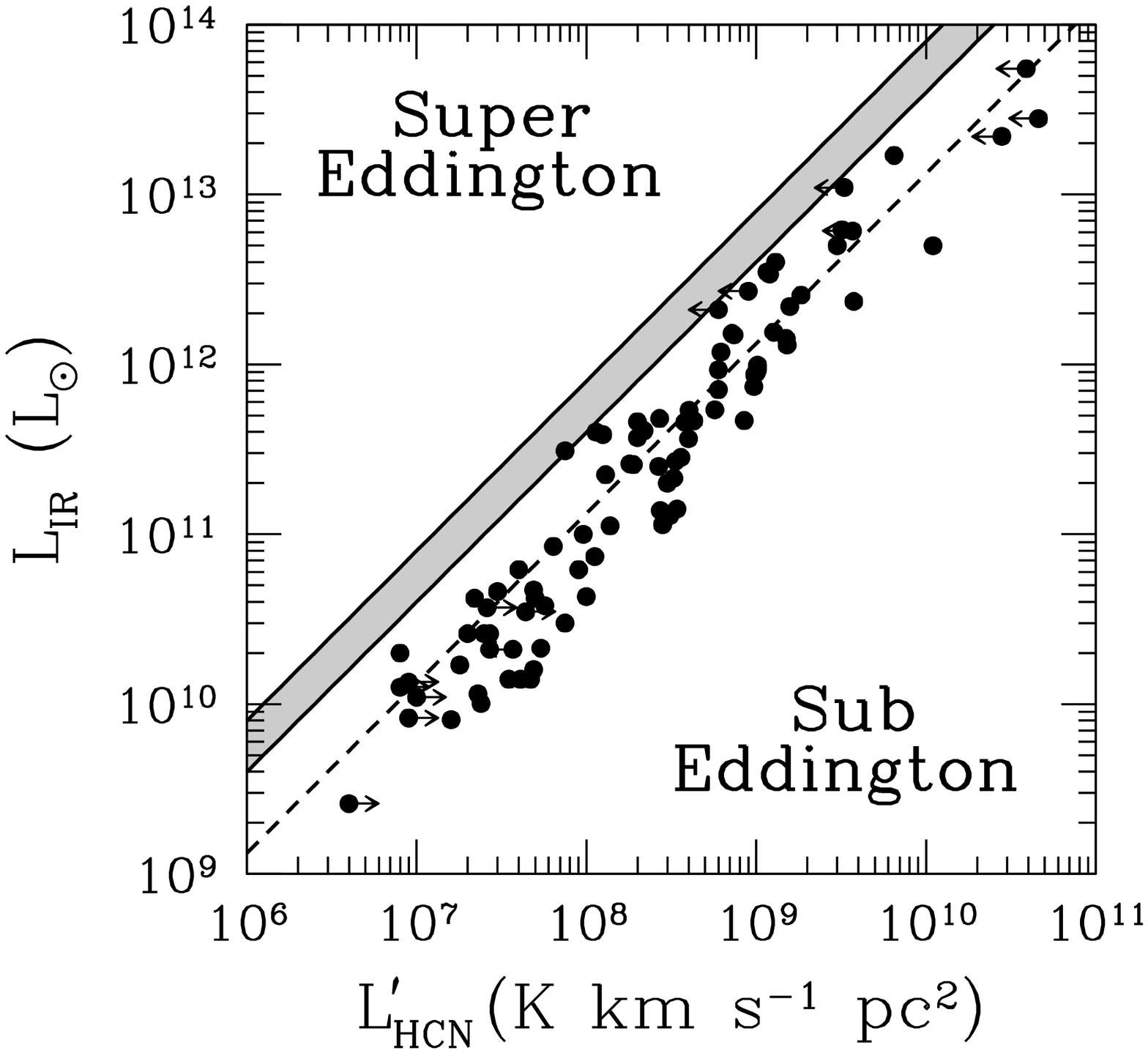}
\vspace{-1cm}
\caption{IR luminosity vs.~HCN line luminosity.  The lines show the optically
thick Eddington limit for our preferred value of the dust opacity (gray region)
and for an enhanced dust-to-gas ratio (dashed line) as is seen in some dusty
starbursts.}
\label{fig:lirlhcn}
\end{center}
\end{minipage}
\hfill
\begin{minipage}[t]{6cm}
\begin{center}
\includegraphics[width=6cm]{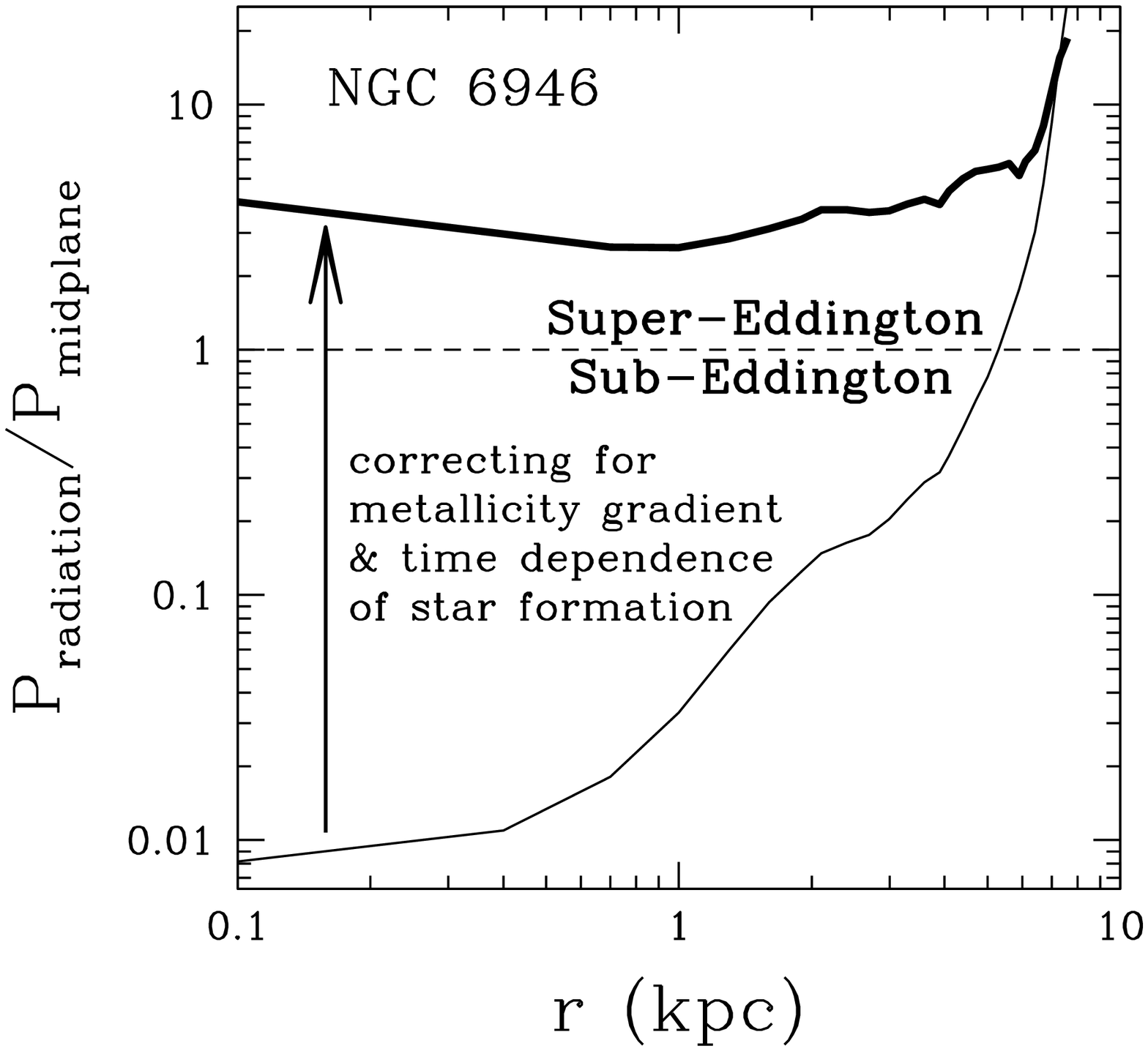}
\vspace{-1cm}
\caption{Eddington ratio ($P_{\rm rad}/P_{\rm mid}$) vs.~radius in NGC 6946
  (thin line).  The arrow shows the effect of correcting the Eddington ratio
  (thick line) for a metallicity gradient and the intermittent nature of
  star-forming disks.}
\label{fig:ngc6946}
\end{center}
\end{minipage}
\end{figure}

$L_{\rm HCN}^\prime$ is proportional to the dense gas mass of galaxies, which is
expected to be optically-thick, and $L_{\rm FIR}$ traces the bolometric
luminosity of star formation.  Thus, radiation pressure feedback predicts a
linear correlation between these two quantities, in good agreement with Figure
\ref{fig:lirlhcn}.  Over $\sim$4 dex in dense gas mass, star-forming galaxies
approach but do not exceed Eddington.  We find similar results for the $L_{\rm
IR}$--$L_{\rm CO}^\prime$ relation and the Schmidt law, but these relations are
complicated by the intermittency of star formation in normal spirals---the
tendency for subregions to dim on a timescale that is short relative to the time
between star-forming events.

Figure \ref{fig:ngc6946} shows the Eddington ratio ($P_{\rm radiation}$/$P_{\rm
midplane}$) as a function of radius in a local spiral (NGC 6946; Leroy \etal \/
\cite{l08}).  At small radii, radiation pressure is significantly subdominant
compared to the midplane pressure required to support the gas disk.  This
discrepancy can be overcome if we account for intermittency and a metallicity
gradient in NGC 6946, since a metallicity gradient would likely increase the
dust-to-gas ratio and decrease the CO-to-H$_2$ conversion factor.  These two
factors, along with the HCN-to-H$_2$ conversion factor, are the primary
observational uncertainties in this analysis, and better constraints on them are
needed to definitively assess radiation pressure on dust as a feedback
mechanism (see AT for details).

\end{document}